\documentclass[11pt]{article}
\usepackage[margin=1in]{geometry}
\usepackage[T1]{fontenc}
\usepackage[utf8]{inputenc}
\usepackage{lmodern,microtype,booktabs,tabularx,array,tikz,graphicx,xcolor}
\usepackage[section]{placeins} 
\usetikzlibrary{arrows.meta,positioning,fit,backgrounds}
\usepackage[hidelinks]{hyperref}

\usepackage[numbers,sort&compress]{natbib}

\title{The Illusion of Evaluation: Challenges in Agentic Product Evaluation Practice}
\author{Abhinav Palia \quad Andrew Gully \quad Drew Brooks\\ \\Microsoft}
\date{}

\begin{document}
\maketitle

\begin{abstract}
Agentic products introduce risks that extend beyond those of their underlying models, arising from their tools, permissions, data access, safeguards, and operating conditions. Existing evaluation instruments provide useful evidence about model behavior and agents in defined harnesses, but their results do not necessarily transfer to deployed products or cover risks created by product configurations and interactions. Identifying and measuring such risks is especially difficult before deployment, when failure pathways, exposure, and outcome distributions may be poorly understood. This raises a fundamental question: what can evaluation evidence establish about the safety of a particular product, and what decisions can it legitimately support?

We call the \emph{illusion of evaluation} the appearance of safety assurance when producing an evaluation result is treated as sufficient evidence for claims that extend beyond what was measured. Drawing on practitioner experience, prior research, and twelve illustrative product contexts, we identify six methodological challenges spanning product-risk identification, ecological and measurement validity, metric interpretation, outcome verification, evidence transfer, and evidentiary sufficiency. We propose a granular transparency-and-decision-relevance record that makes risks, system configurations, assumptions, uncertainty, and evidentiary limits explicit. The proposal is a reporting and reasoning aid, not proof of safety or a deployment-approval standard. We conclude with open questions on valid product-level measurements, evidence transfer, and decision sufficiency.
\end{abstract}

\section{Introduction}

Evaluations have become central to assessing generative AI systems. Behavioral benchmarks, adversarial tests, and agent environments examine model behavior, task completion, and selected failure modes \citep{zhou2024webarena,jimenez2024swebench,yao2024taubench,debenedetti2024agentdojo}. Some test an isolated model; others include agents, tools, simulated users, policies, or functional environments. All nevertheless produce evidence for specified systems and conditions. The existence of an instrument does not establish its suitability for evaluating a particular product.

This distinction matters because products extend models through retrieval, persistent state, external tools, permissions, safeguards, and autonomous actions. A product may inherit a model risk, but it does not automatically inherit the safety conclusion from a model evaluation. Product design can change the manifestation and consequence of that risk, introduce new failure pathways, or prevent a model failure from becoming a consequential outcome. For example, a security assistant may accurately summarize the scans it ran while failing to examine critical assets, and a coding agent may complete a requested change while modifying files outside its authorized scope. In each case, a plausible output can coexist with an unsafe product outcome.

Product evaluation is difficult for two related reasons. First, the instrument must validly operationalize a consequential product risk under sufficiently realistic conditions. Second, the resulting evidence must justify the scope of the claim and the decision assigned to it. Controlled benchmarks can be reproducible yet ecologically distant from the target workflow. A detected failure may establish a vulnerability without estimating its prevalence. Conversely, observing few failures cannot establish low deployment risk across untested pathways. These are questions of measurement suitability, ecological validity, interpretation, and inference.

Evaluations also support different decisions: discovery and remediation; selection between configurations; validation of a mitigation; definition of permissions or operating restrictions; quality--safety trade-offs; and judgments about suitability for intended users and environments. The evidence required differs. One verified consequential failure may justify remediation. Comparative selection requires matched conditions. A deployment decision may require product-specific outcome evidence, controls, exposure assumptions, and monitoring. An undifferentiated ``safety score'' obscures these differences.

We use \emph{illusion of evaluation} for the appearance of assurance created when an evaluation result is taken to justify a claim beyond its evidentiary scope. We contribute (1) a product-centered account of six recurring breaks in the chain from risk to decision, checked against twelve illustrative contexts; and (2) a granular Transparency and Decision-Relevance Record (TDRR) that separates observations, assumptions, transfer arguments, uncertainty, and unsupported inferences. The TDRR does not replace valid measurement, safety cases, expert judgment, or empirical evidence about outcomes. It makes the boundaries among them inspectable.

\section{Related Work and Positioning}

\paragraph{Measurement and benchmark validity.}
Work on construct validity, benchmark design, and evaluation comparison distinguishes a measured construct from the broader property attributed to a system \citep{wallach2025measurement,raji2021ai,chouldechova2025comparison,eriksson2025trust}. Reproducibility or judge agreement is not enough: the operationalization, test population, and interpretation must correspond to the intended claim. Product evaluation adds the problem of representing consequential outcomes within a configured sociotechnical system.
\paragraph{Model and agent evaluation.}
WebArena, SWE-bench, and $\tau$-bench evaluate behavior in specified tasks and environments; AgentDojo evaluates prompt-injection attacks and defenses for tool-using agents \citep{zhou2024webarena,jimenez2024swebench,yao2024taubench,debenedetti2024agentdojo}. These instruments go beyond isolated model prompts, but their tasks, simulated users, tools, permissions, policies, and outcome criteria need not match a target product. Their ecological validity and the transferability of findings must therefore be established, not assumed. Specification-driven approaches such as ASSERT expose how rubrics, scenarios, dialogue conditions, judges, and evidence thresholds produce a measurement \citep{fogliato2026assert}. This transparency is necessary, but scenario generation and explicit specifications do not by themselves establish product-risk coverage or decision sufficiency. 

\paragraph{Auditing and safety arguments.}
Internal auditing emphasizes system assessment throughout development \citep{raji2020accountability}; safety-case research organizes claims, evidence, assumptions, and arguments \citep{clymer2024safetycases}. Our proposal is complementary. It neither supplies a complete assurance argument nor validates a product. It asks whether a specific measurement is relevant to a specific product claim and decision.

\section{Basis and Scope}

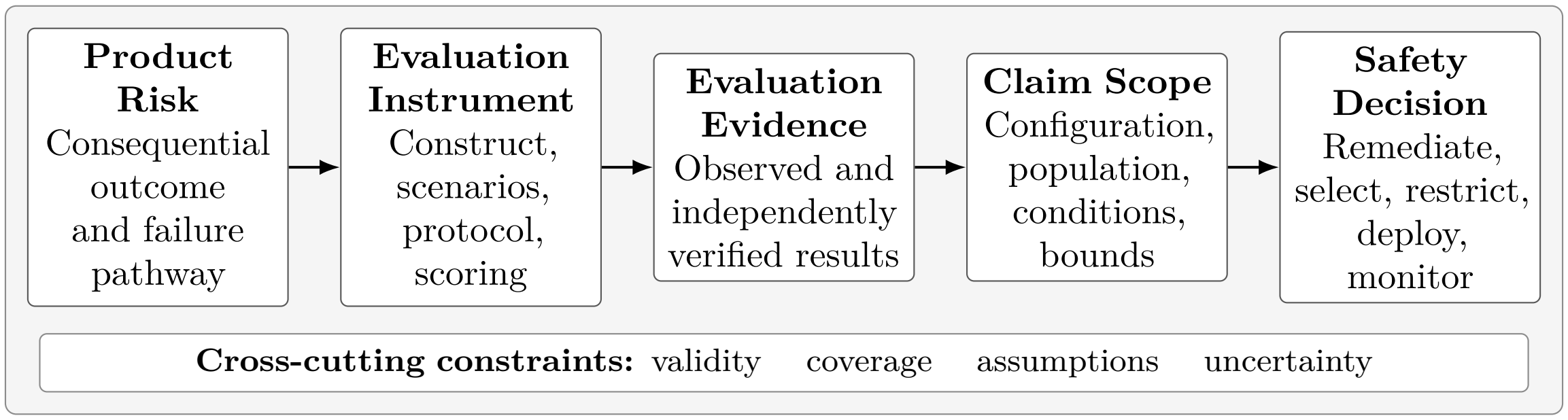
\begin{figure}[h]
\centering
\begin{tikzpicture}[
  font=\small,
  node distance=5mm,
  stage/.style={draw=black!65,rounded corners=2pt,align=center,text width=2.25cm,minimum height=1.25cm,inner sep=4pt,fill=white},
  arr/.style={-{Latex[length=2mm]},line width=.75pt},
  band/.style={draw=black!45,rounded corners=3pt,fill=black!4,inner sep=6pt},
  label/.style={font=\footnotesize,align=center}
]
\node[stage] (risk) {\textbf{Product Risk}\\Consequential outcome and failure pathway};
\node[stage,right=of risk] (instrument) {\textbf{Evaluation Instrument}\\Construct, scenarios, protocol, scoring};
\node[stage,right=of instrument] (evidence) {\textbf{Evaluation Evidence}\\Observed and independently verified results};
\node[stage,right=of evidence] (claim) {\textbf{Claim Scope}\\Configuration, population, conditions, bounds};
\node[stage,right=of claim] (decision) {\textbf{Safety Decision}\\Remediate, select, restrict, deploy, monitor};
\draw[arr] (risk) -- (instrument);
\draw[arr] (instrument) -- (evidence);
\draw[arr] (evidence) -- (claim);
\draw[arr] (claim) -- (decision);
\node[label,draw=black!45,rounded corners=2pt,fill=white,below=5mm of evidence,text width=.86\textwidth,inner sep=4pt] (cross) {\textbf{Cross-cutting constraints:} validity \quad coverage \quad assumptions \quad uncertainty};
\begin{scope}[on background layer]
  \node[band,fit=(risk)(instrument)(evidence)(claim)(decision)(cross)] {};
\end{scope}
\end{tikzpicture}
\caption{A decision-oriented inferential chain. Each arrow requires an argument. Validity, coverage, assumptions, and uncertainty cut across all stages; a break cannot be repaired merely by producing a score.}
\label{fig:chain}
\end{figure}

This is a practitioner-informed position and conceptual synthesis, not a systematic empirical study. The six challenges draw on experience with generative AI evaluation and recurring issues in the evaluation and assurance literature. They follow the path from product-risk identification through instrument design, evidence production, claim definition, and decision-making (Figure~\ref{fig:chain}). They are a problem characterization, not an exhaustive or empirically validated taxonomy.

We use twelve contexts: workplace assistance, coding, clinical documentation, data analytics, security operations, human resources, customer support, financial research, shopping, browser automation, legal review, and autonomous research (Table~\ref{tab:cases}). They vary in data sensitivity, tools and permissions, autonomy and oversight, affected stakeholders, and the severity, observability, and reversibility of failures. They are analytical prompts selected for breadth, not a representative sample or a dataset of observed incidents.

The contexts separate model risk from evidence transfer. Harmful-content generation, susceptibility to manipulation, or unreliable reasoning may originate partly in a model. Whether model-level evidence transfers depends on the product's instructions, interfaces, tools, access controls, safeguards, users, and consequences. Separately, the product may introduce risks---such as unauthorized transactions or incomplete workflow execution---that the model evaluation never measured.

\section{Challenges in Product Evaluation}

\paragraph{1. Identifying product-specific consequential risks.}
Existing taxonomies organize a broad range of ethical, social, safety, and misuse risks associated with language models and AI systems \citep{taxonomy1,SLATTERY2026101517}. These taxonomies provide useful starting points, but model-centered or generic categories
can miss harms created by a product's authorized workflows, data access, tools, and component interactions. In clinical documentation, fluent transcription may still place an incorrect dosage or omitted negation in the saved record. In security operations, a correct summary of completed scans may conceal assets that were never scanned. The relevant risk concerns the consequential state of the workflow, not only the generated text. Risk analysis and red teaming can discover pathways, but neither guarantees coverage. Evaluation begins by specifying the stakeholder or asset, adverse outcome, entry point, trust and authorization boundaries, and the pathway by which the configured
product can cause or enable the outcome.

\paragraph{2. Establishing ecological validity and measurement suitability.}
After a risk is specified, an instrument must operationalize it under conditions that correspond to the intended claim. A customer-support judge may reward a completed refund without checking account control, policy compliance, or the resulting ledger state. A simulated tool may omit access-control behavior that dominates the real product. Predeployment tests face a particular difficulty: realistic users, exposure, drift, and rare interactions may not yet be observable. Even after deployment, privacy obligations and restrictions on retaining or inspecting actual user interactions can limit access to the data needed to characterize real workflows and failure conditions. Privacy-preserving alternatives---including consented studies, de-identified traces, aggregate or differentially private telemetry, protected analysis environments, and synthetic interactions---can reduce these constraints \citep{dwork2014algorithmic,garfinkel2023deidentifying}. However, privacy transformations can reduce data utility or remove contextual and distributional properties needed for ecologically valid evaluation, while synthetic data may provide unpredictable privacy--utility trade-offs \citep{stadler2022synthetic}. Benchmark realism is not binary; its relevant dimensions include workflow fidelity, data and user populations, permissions, incentives, time horizons, and consequences. Instrument suitability should therefore be argued risk-by-risk and claim-by-claim. Evaluation awareness may further complicate generalization \citep{needham2025largelanguagemodelsknow}.

\paragraph{3. Interpreting metrics: discovery is not prevalence.}
Adversarial and generated scenarios are often optimized to find failures. A verified failure can establish possibility and motivate remediation without estimating how often it will occur. Prevalence claims instead require a defensible target population, sampling frame, exposure model, denominator, and treatment of clustering or dependence. The same observed rate can imply different risks when scenario severity, opportunity for exposure, or outcome definitions differ. Comparisons require aligned populations and procedures \citep{chouldechova2025comparison}. Reports should state whether a metric supports discovery, regression detection, matched comparison, mitigation testing, prevalence estimation, or another use. It should not silently migrate between these purposes.

\paragraph{4. Verifying consequential outcomes.}
A model statement, tool request, tool response, state change, and authorization judgment constitute distinct forms of evidence. An agent saying that a refund was issued does not prove that the transaction occurred, while a tool-call trace establishes only the recorded request and response. Verifying the consequence requires observing the relevant environment state---for example, comparing the final database state with an annotated goal state, as in $\tau$-bench, or executing tests against the modified repository, as in SWE-bench \citep{yao2024taubench,jimenez2024swebench}. Whether the action was authorized requires an additional comparison with the user's identity and request, the agent's permissions, and applicable policy; benchmarks such as AgentDojo similarly use task- and environment-specific checks to distinguish utility from security violations \citep{debenedetti2024agentdojo}. Such verification is inherently product-dependent: a valid evaluator must understand the product's state model, authorization semantics, tool behavior, persistence, and external side effects. Developing product-level evaluations may therefore require substantial additional infrastructure---including instrumented tools, state snapshots, authorization oracles, audit-log correlation, controlled test environments, and domain-specific outcome assertions---rather than a generic model or trace evaluator. Verification must ultimately match the consequence: message content, recipient, and delivery state for email; changed files, test behavior, and repository scope for coding; or record identity and field values for clinical documentation.

\paragraph{5. Separating model risk from evidence transfer.}
A product may inherit a model's tendency without inheriting either its measured rate or the safety conclusion attached to that rate. Conversely, system controls may block a harmful model behavior from producing harm. Evidence transfer is the argument that a result from a source configuration remains informative for a target claim. It requires identifying matched and changed components, mechanisms affected by those changes, and revalidation triggers. Model-level prompt-injection evidence, for example, does not alone establish confidentiality for an assistant with enterprise retrieval and email permissions; product access controls may also render some model failures non-consequential. Risk inheritance, mitigation effectiveness, and evidence transfer should therefore be reported separately. Post-training interventions can also be sensitive to subsequent changes \citep{qi2024finetuning,wei2024brittleness}.

\paragraph{6. Determining decision-specific evidentiary sufficiency.}
Evidence adequate for one decision may be insufficient for another. A single credible high-impact failure can justify remediation even when its prevalence is unknown. For example, one verified instance in which a model provides actionable assistance for constructing a chemical weapon may be sufficient to trigger mitigation or restrict access; a measured failure rate is less useful to establish that the capability exists and could have severe consequences. Selecting between two configurations, by contrast, requires comparable measurements, while restoring a mitigation after a regression requires evidence tied to the relevant causal pathway. 

Adversarial evaluations create an additional difficulty: the observed failure rate depends on the attacker's knowledge, capabilities, strategy, and interaction budget \citep{vassilev2025adversarial,anil2024manyshot}. Because adaptive adversaries can search across prompts, contexts, tools, and multi-turn strategies, a finite test set cannot establish coverage of all plausible attacks; red-teaming results should therefore be interpreted as evidence about the tested search process rather than an exhaustive characterization of what an adversary could elicit \citep{perez2022redteaming,ganguli2022redteaming}. Deployment for an intended audience may consequently require product-specific outcome tests, architectural controls, expert review, exposure assumptions, monitoring, and response capacity. No universal passing score resolves these differences. The decision-maker must state the proposed action, the claim necessary to justify it, the adversary model and test budget, acceptable residual uncertainty, and which assumptions or attack pathways remain outside the evaluation. Frameworks such as the NIST AI RMF provide broader risk-management context \citep{nist2023airmf}, but they do not determine the evidentiary sufficiency of a particular score.

\section{Reporting Evidentiary Boundaries}
\label{sec:reporting}

An evaluation report should explain how a result was produced, what it supports, and why it matters for a decision. Model cards, internal auditing, measurement specifications, and safety cases show the value of documenting intended uses, conditions, evidence, assumptions, and limitations \citep{mitchell2019modelcards,raji2020accountability,clymer2024safetycases,fogliato2026assert}. Documentation alone, however, cannot make an unsuitable instrument valid or an unsupported inference sound.

We propose the \emph{Transparency and Decision-Relevance Record} (TDRR), a claim-level procedure that links one versioned product configuration and one bounded claim to one decision. It has three functions: disclose how the evidence was produced; interpret what the evidence supports and leaves unresolved; and state how those limits affect the proposed action. The complete field-level template appears in Appendix~\ref{app:tdrr}.

\subsection{Interpretation and Decision Consequences}

For each material claim, reviewers classify the evidence as \emph{supporting}, \emph{partially supporting}, \emph{not supporting}, \emph{contradicting}, or \emph{indeterminate}. The judgment should consider construct validity, scenario coverage, ecological validity, outcome verification, configuration match, metric purpose, and uncertainty. It should yield the narrowest supported claim, not merely a score relative to a generic threshold. No observed failures in a finite test set is not evidence of absence without a justified population, sampling procedure, and exposure model. Conversely, one verified high-severity failure may contradict a claim or justify remediation without a prevalence estimate.

Evidentiary gaps matter only when their consequences for the proposed action are explicit. Table~\ref{tab:decisionrules} gives illustrative mappings; they are not universal decision rules.

\begin{table}[t]
\centering
\small
\caption{Illustrative relationships between evidentiary findings and decision consequences. Appropriate actions remain product-, risk-, and context-dependent.}
\label{tab:decisionrules}
\begin{tabularx}{\textwidth}{@{}p{0.38\textwidth}X@{}}
\toprule
\textbf{Evidentiary finding} & \textbf{Possible decision consequence}\\
\midrule

Verified high-severity failure &
Remediate the failure pathway, restrict the affected capability, strengthen controls, or defer deployment. A prevalence estimate may not be necessary to justify action.\\

Instrument does not measure the consequential risk &
Do not use the score to support the product claim. Redesign the evaluation or narrow the claim to the construct actually measured.\\

Evidence supports discovery but not prevalence &
Treat verified failures as actionable findings, but do not interpret the observed rate as deployment frequency or residual risk.\\

Consequential outcome was not independently verified &
Limit the conclusion to generated behavior, attempted action, or recorded tool interaction. Add product-specific state verification before making an outcome claim.\\

Source and target configurations differ materially &
Provide a mechanism-based transfer argument, revalidate the changed components, restrict the claim to the source configuration, or treat the result as unsupported for the target product.\\

Important pathways or user populations remain untested &
Narrow the intended use or claim; introduce compensating controls and monitoring; obtain additional evidence; or defer the decision.\\

Evidence contradicts the required claim &
Do not approve the decision on the basis of that claim unless the failure is remediated, the claim is revised, or an alternative argument and controls address the consequence.\\

Residual uncertainty is accepted &
Name the accountable decision-maker; document why the uncertainty is tolerable; specify restrictions, monitoring, incident response, expiration, and revalidation triggers.\\

Evidence is indeterminate &
Do not translate the result into a positive safety claim. Resolve methodological defects or record ``insufficient evidence'' as the decision outcome.\\

\bottomrule
\end{tabularx}
\end{table}

The same result can therefore support remediation but not a prevalence estimate, or selection between configurations but not deployment of either. The TDRR does not prescribe a universal risk threshold. It requires the applicable decision rule, its relevance to the harm and context, and the person authorized to accept residual uncertainty. Where no defensible threshold exists, the record should not manufacture one from a score.

\subsection{Use and Review}

The TDRR should be applied prospectively and revisited as the product, evidence, or decision changes:

\begin{enumerate}
    \item \textbf{Begin with the decision.} State the proposed action, alternatives, and the exact claim required to justify it.

    \item \textbf{Specify the risk and boundary.} Record the consequential outcome, affected stakeholders, failure pathway, target configuration, users, and operating conditions. Separate model risk from product-created pathways.

    \item \textbf{Define and produce the evidence.} State whether the purpose is discovery, comparison, mitigation validation, prevalence estimation, or control verification. Design the construct, scenarios, protocol, scoring, and outcome verification accordingly; version the configuration and preserve auditable artifacts.

    \item \textbf{Interpret and bound.} Classify the evidence, state the narrowest supported claim, identify unsupported inferences, and document assumptions required for transfer or generalization.

    \item \textbf{Challenge the argument.} An independent or cross-functional reviewer should test whether the evidence supports the interpretation. Preserve material disagreements and translate gaps into additional evidence, controls, restrictions, monitoring, escalation, or deferral.

    \item \textbf{Decide and revalidate.} Record the action, accountable decision-maker, accepted uncertainty, restrictions, monitoring, and triggers for review when the system or environment changes.
\end{enumerate}

Every record should conclude with a compact synthesis:

\begin{quote}
For product configuration \emph{C}, evidence \emph{E} \textbf{[supports / partially supports / does not support / contradicts / leaves indeterminate]} claim \emph{K} under conditions \emph{A}. The evidence does not establish \emph{U}. Because decision \emph{D} requires \emph{K} and depends on residual uncertainty \emph{U}, the recommended action is \emph{R}, subject to restrictions, monitoring, and revalidation triggers \emph{T}.
\end{quote}

This synthesis prevents a collection of details from substituting for an interpretation and identifies what would have to change to alter the decision.

\subsection{Safeguards and Boundaries}

The TDRR could become compliance theater if completion is rewarded rather than defensible reasoning. Material fields should link to inspectable artifacts; claims about coverage, representativeness, transfer, and mitigation should require justification; and ``not assessed,'' ``withheld,'' and ``insufficient evidence'' must remain permissible outcomes. Consequential decisions should preserve reviewer disagreement, name who accepted residual uncertainty, and expire when specified aspects of the product or environment change.

The TDRR improves inspectability, not validity by documentation. It cannot make an unsuitable instrument suitable, guarantee coverage, infer prevalence from discovery-oriented tests, set acceptable residual risk, or prove safety. Nor does it replace domain expertise, user research, privacy analysis, controls, monitoring, incident response, or a complete safety argument. Reporting depth should be proportionate to the severity, reversibility, and uncertainty of the decision.

Transparency also has legitimate privacy, security, and evaluation-integrity limits. Public reports can disclose constructs, methods, aggregate strata, versions, exclusions, verification procedures, uncertainty, and the nature of withheld material while sensitive artifacts remain available to authorized reviewers. Withholding should be justified and its effect on reproducibility or scrutiny stated.

Within these limits, transparency reduces pressure on a score to serve as a definitive safety judgment. Separating what was observed from what is assumed, transferred, untested, or unknown helps expose evaluation theater and reduces the risk that incomplete evidence will misguide decision-makers. This is a responsible-AI practice, not validation of the evidence. The TDRR should remain a proposed method until studies show whether it narrows claims, reveals unsupported transfers, changes tests or controls, improves reviewer agreement, or affects decisions.
\section{Research Agenda and Conclusion}

The six challenges motivate three connected research directions. First, how can consequential product risks be identified and operationalized without reducing them to generic model behaviors? Work is needed on coverage arguments, realistic predeployment populations, and outcome verification across domains. Second, when are measurements ecologically valid, comparable, and transferable across configurations? This requires empirical study of which product changes alter failure mechanisms and how transfer assumptions should be tested. Third, what constitutes sufficient evidence for different decisions? Studies should examine how evaluation findings, controls, uncertainty, monitoring, and stakeholder values combine in remediation, restriction, selection, and deployment judgments.

The illusion of evaluation arises when the existence of a result is mistaken for sufficient evidence of product safety. A valid measurement can be useful yet insufficient for a broader product claim; a failure discovered in an adversarial test can be consequential without estimating prevalence. By making the full chain---Product Risk, Evaluation Instrument, Evaluation Evidence, Claim Scope, and Safety Decision---explicit, the TDRR aims to expose where evidence ends and assumption begins. Whether it improves evaluation practice is an empirical question, and should be treated as such.

\appendix
\section*{Appendix}
\section{Illustrative Product Contexts}

\begin{table}[h]
\centering\small
\caption{Twelve illustrative product contexts and possible consequential risks. These are analytical examples, not observed incidents or a systematic sample.}
\label{tab:cases}
\begin{tabularx}{\textwidth}{@{}p{0.23\textwidth}X@{}}
\toprule
\textbf{Product context} & \textbf{Possible consequential risk}\\
\midrule
Workplace assistant & Confidential disclosure or unauthorized email\\
Coding agent & Insecure code, unauthorized edits, or secret exposure\\
Clinical documentation & Incorrect dosage, omitted negation, or wrong patient record\\
Data analytics agent & Unauthorized database access or destructive queries\\
Security operations & Missed assets, false assurance, or disruptive remediation\\
HR assistant & Sensitive disclosure or mistaken identity\\
Customer support & Unauthorized account change or incorrect refund\\
Financial research & Unsupported analysis or stale evidence\\
Shopping assistant & Misleading claims or unauthorized purchases\\
Browser automation & Unauthorized transaction or destructive action\\
Legal review & Missed obligations or privileged disclosure\\
Autonomous research & Unsupported conclusions or unsafe tool actions\\
\bottomrule
\end{tabularx}
\end{table}

\section{Transparency and Decision-Relevance Record}
\label{app:tdrr}

Each field should contain a value, a stable reference to an artifact, ``not assessed,'' or ``withheld'' with a reason and its implications for review. Text alone is not evidence: material assertions should be tied to evaluation artifacts, system documentation, empirical evidence, or an identified expert judgment.

\begin{table}[h]
\centering
\footnotesize
\caption{Detailed TDRR template for a product-level safety claim. Completion is not itself evidence of safety.}
\label{tab:tdrr}
\setlength{\tabcolsep}{4pt}
\renewcommand{\arraystretch}{1.1}
\begin{tabularx}{\textwidth}{@{}p{0.22\textwidth}X@{}}
\toprule
\textbf{Category} & \textbf{Required fields and judgments}\\
\midrule
1. Proposed decision & Action, alternatives, decision-maker, risk owner, intended users and setting, deadline, and review date.\\
2. Required claim & Exact wording, claim type, target configuration and population, conditions, time horizon, exceptions, and relevance to the decision.\\
3. System boundary & Model and version, prompts, orchestration, tools, permissions, data sources, safeguards, human approvals, autonomy, dependencies, and build identifiers.\\
4. Risk specification & Stakeholder or asset, adverse outcome, severity, reversibility, failure pathway, actor, entry point, trust boundary, authorization rule, exposure assumptions, and exclusions.\\
5. Measurement construct & Target property, operational definition, unit of analysis, rubric, reference and boundary cases, and relationship to the consequential outcome.\\
6. Scenario population & Source, target population if claimed, sampling or generation, inclusion and exclusion, strata, interaction budget, deduplication, coverage rationale, blind spots, and metric purpose.\\
7. Execution protocol & Roles, participants or simulators, horizon, live or mocked tools, retries, stop rules, adversary knowledge and adaptation, controls, and differences from deployment.\\
8. Scoring and adjudication & Success definition, judge and version, instructions, evidence threshold, validation, human review, disagreements, adjudication, and known scoring errors.\\
9. Outcome verification & Model statement, requested action, tool response, resulting state, persistence, authorization check, independent assertions or logs, and product-specific instrumentation.\\
10. Results and uncertainty & Numerator, denominator, sample and strata, missing runs, justified intervals, severity, failure taxonomy, sensitivity analyses, and comparison conditions.\\
11. Controls and residual risk & Controls exercised, safeguards, bypass conditions, defense-in-depth, utility costs, untested pathways, monitoring, response, and remaining risks.\\
12. Evidence transfer & Source and target configurations, matched and changed components, transfer mechanism, supporting evidence, assumptions, limits, and revalidation triggers.\\
13. Decision synthesis & Evidence classification, narrowest supported claim, unsupported inferences, assumptions, residual uncertainty, consequences, restrictions, additional evidence, reviewers, disagreements, rationale, triggers, artifacts, and revision history.\\
\bottomrule
\end{tabularx}
\end{table}

\clearpage

\bibliographystyle{unsrtnat}
\bibliography{references}

@inproceedings{zhou2024webarena,
  title={WebArena: A Realistic Web Environment for Building Autonomous Agents},
  author={Zhou, Shuyan and Xu, Frank F. and Zhu, Hao and Zhou, Xuhui and Lo, Robert and Sridhar, Abishek and Cheng, Xianyi and Ou, Tianyue and Bisk, Yonatan and Fried, Daniel and Alon, Uri and Neubig, Graham},
  booktitle={The Twelfth International Conference on Learning Representations}, year={2024},
  url={https://openreview.net/forum?id=oKn9c6ytLx}}

@inproceedings{jimenez2024swebench,
  title={{SWE}-bench: Can Language Models Resolve Real-World {GitHub} Issues?},
  author={Jimenez, Carlos E. and Yang, John and Wettig, Alexander and Yao, Shunyu and Pei, Kexin and Press, Ofir and Narasimhan, Karthik R.},
  booktitle={The Twelfth International Conference on Learning Representations}, year={2024},
  url={https://openreview.net/forum?id=VTF8yNQM66}}

@misc{yao2024taubench,
  title={{${\tau}$-bench}: A Benchmark for Tool-Agent-User Interaction in Real-World Domains},
  author={Yao, Shunyu and Shinn, Noah and Razavi, Pedram and Narasimhan, Karthik}, year={2024},
  eprint={2406.12045}, archivePrefix={arXiv}, primaryClass={cs.AI},
  doi={10.48550/arXiv.2406.12045}, url={https://arxiv.org/abs/2406.12045}}

@inproceedings{debenedetti2024agentdojo,
  title={AgentDojo: A Dynamic Environment to Evaluate Prompt Injection Attacks and Defenses for {LLM} Agents},
  author={Debenedetti, Edoardo and Zhang, Jie and Balunovic, Mislav and Beurer-Kellner, Luca and Fischer, Marc and Tram{\`e}r, Florian},
  booktitle={Advances in Neural Information Processing Systems 37: Datasets and Benchmarks Track},
  year={2024}, url={https://openreview.net/forum?id=m1YYAQjO3w}}

@inproceedings{raji2020accountability,
  title={Closing the {AI} Accountability Gap: Defining an End-to-End Framework for Internal Algorithmic Auditing},
  author={Raji, Inioluwa Deborah and Smart, Andrew and White, Rebecca N. and Mitchell, Margaret and Gebru, Timnit and Hutchinson, Ben and Smith-Loud, Jamila and Theron, Daniel and Barnes, Parker},
  booktitle={Proceedings of the 2020 Conference on Fairness, Accountability, and Transparency},
  pages={33--44}, publisher={Association for Computing Machinery}, year={2020},
  doi={10.1145/3351095.3372873}}

@techreport{nist2023airmf,
  title={Artificial Intelligence Risk Management Framework ({AI RMF 1.0})},
  author={{National Institute of Standards and Technology}},
  institution={National Institute of Standards and Technology}, number={NIST AI 100-1},
  year={2023}, doi={10.6028/NIST.AI.100-1}}

@inproceedings{qi2024finetuning,
  title={Fine-Tuning Aligned Language Models Compromises Safety, Even When Users Do Not Intend To!},
  author={Qi, Xiangyu and Zeng, Yi and Xie, Tinghao and Chen, Pin-Yu and Jia, Ruoxi and Mittal, Prateek and Henderson, Peter},
  booktitle={The Twelfth International Conference on Learning Representations}, year={2024},
  url={https://openreview.net/forum?id=hTEGyKf0dZ}}

@inproceedings{wei2024brittleness,
  title={Assessing the Brittleness of Safety Alignment via Pruning and Low-Rank Modifications},
  author={Wei, Boyi and Huang, Kaixuan and Huang, Yangsibo and Xie, Tinghao and Qi, Xiangyu and Xia, Mengzhou and Mittal, Prateek and Wang, Mengdi and Henderson, Peter},
  booktitle={Proceedings of the 41st International Conference on Machine Learning},
  series={Proceedings of Machine Learning Research}, volume={235}, publisher={PMLR}, year={2024},
  url={https://proceedings.mlr.press/v235/wei24f.html}}

@misc{wallach2025measurement,
  title={Position: Evaluating Generative {AI} Systems Is a Social Science Measurement Challenge},
  author={Wallach, Hanna and Desai, Meera and Cooper, A. Feder and Wang, Angelina and Atalla, Chad and Barocas, Solon and Blodgett, Su Lin and Chouldechova, Alexandra and Corvi, Emily and Dow, P. Alex and Garcia-Gathright, Jean and Olteanu, Alexandra and Pangakis, Nicholas and Reed, Stefanie and Sheng, Emily and Vann, Dan and Vaughan, Jennifer Wortman and Vogel, Matthew and Washington, Hannah and Jacobs, Abigail Z.},
  year={2025}, eprint={2502.00561}, archivePrefix={arXiv}, primaryClass={cs.CY},
  doi={10.48550/arXiv.2502.00561}, url={https://arxiv.org/abs/2502.00561}}

@misc{raji2021ai,
  title={{AI} and the Everything in the Whole Wide World Benchmark},
  author={Raji, Inioluwa Deborah and Bender, Emily M. and Paullada, Amandalynne and Denton, Emily and Hanna, Alex},
  year={2021}, eprint={2111.15366}, archivePrefix={arXiv}, primaryClass={cs.LG},
  doi={10.48550/arXiv.2111.15366}, url={https://arxiv.org/abs/2111.15366}}

@misc{chouldechova2025comparison,
  title={Comparison Requires Valid Measurement: Rethinking Attack Success Rate Comparisons in {AI} Red Teaming},
  author={Chouldechova, Alexandra and Cooper, A. Feder and Barocas, Solon and Palia, Abhinav and Vann, Dan and Wallach, Hanna},
  year={2026}, eprint={2601.18076}, archivePrefix={arXiv}, primaryClass={cs.LG},
  doi={10.48550/arXiv.2601.18076}, url={https://arxiv.org/abs/2601.18076}}

@misc{eriksson2025trust,
  title={Can We Trust {AI} Benchmarks? An Interdisciplinary Review of Current Issues in {AI} Evaluation},
  author={Eriksson, Maria and Purificato, Erasmo and Noroozian, Arman and Vinagre, Jo{\~a}o and Chaslot, Guillaume and G{\'o}mez, Emilia and Fern{\'a}ndez-Llorca, David},
  year={2025}, eprint={2502.06559}, archivePrefix={arXiv}, primaryClass={cs.AI},
  doi={10.48550/arXiv.2502.06559}, url={https://arxiv.org/abs/2502.06559}}

@inproceedings{fogliato2026assert,
  title={{ASSERT}: A Measurement Pipeline for {GenAI} Audits},
  author={Fogliato, Riccardo and Palia, Abhinav and Wang, Xiawei and Sheng, Emily and Atalla, Chad and Garcia-Gathright, Jean and Pangakis, Nicholas and Tan, Sharman and Vann, Dan and Washington, Hannah and Dow, P. Alex and Elfardy, Heba and Wallach, Hanna and Atluri, Sandeep},
  editor={Atanasova, Pepa and Haffari, Gholamreza and Wang, Minghan},
  booktitle={Proceedings of the 2026 Conference on Empirical Methods in Natural Language Processing: Industry Track},
  pages={681--698}, address={Budapest, Hungary}, publisher={Association for Computational Linguistics},
  month=oct, year={2026}, doi={10.18653/2026.emnlp-industry.41},
  url={https://aclanthology.org/2026.emnlp-industry.41/}}

@misc{needham2025largelanguagemodelsknow,
  title={Large Language Models Often Know When They Are Being Evaluated},
  author={Needham, Joe and Edkins, Giles and Pimpale, Govind and Bartsch, Henning and Hobbhahn, Marius},
  year={2025}, eprint={2505.23836}, archivePrefix={arXiv}, primaryClass={cs.CL},
  doi={10.48550/arXiv.2505.23836}, url={https://arxiv.org/abs/2505.23836}}

@misc{clymer2024safetycases,
  title={Safety Cases: How to Justify the Safety of Advanced {AI} Systems},
  author={Clymer, Joshua and Gabrieli, Nick and Krueger, David and Larsen, Thomas},
  year={2024}, eprint={2403.10462}, archivePrefix={arXiv}, primaryClass={cs.CY},
  doi={10.48550/arXiv.2403.10462}, url={https://arxiv.org/abs/2403.10462}}

@inproceedings{taxonomy1,
  title={Taxonomy of Risks Posed by Language Models},
  author={Weidinger, Laura and Uesato, Jonathan and Rauh, Maribeth and Griffin, Conor and Huang, Po-Sen and Mellor, John and Glaese, Amelia and Cheng, Myra and Balle, Borja and Kasirzadeh, Atoosa and Biles, Courtney and Brown, Sasha and Kenton, Zac and Hawkins, Will and Stepleton, Tom and Birhane, Abeba and Hendricks, Lisa Anne and Rimell, Laura and Isaac, William and Haas, Julia and Legassick, Sean and Irving, Geoffrey and Gabriel, Iason},
  booktitle={Proceedings of the 2022 ACM Conference on Fairness, Accountability, and Transparency},
  pages={214--229}, publisher={Association for Computing Machinery}, address={New York, NY, USA},
  year={2022}, doi={10.1145/3531146.3533088}}

@article{SLATTERY2026101517,
  title={The {AI} Risk Repository: A Meta-Review, Database, and Taxonomy of Risks from Artificial Intelligence},
  author={Slattery, Peter and Saeri, Alexander K. and Grundy, Emily A. C. and Graham, Jess and Noetel, Michael and Uuk, Risto and Dao, James and Pour, Soroush and Casper, Stephen and Thompson, Neil},
  journal={Patterns}, volume={7}, number={5}, pages={101517}, year={2026},
  doi={10.1016/j.patter.2026.101517}}

@book{dwork2014algorithmic,
  title={The Algorithmic Foundations of Differential Privacy}, author={Dwork, Cynthia and Roth, Aaron},
  series={Foundations and Trends in Theoretical Computer Science}, volume={9},
  pages={211--407}, publisher={Now Publishers}, year={2014}, doi={10.1561/0400000042}}

@techreport{garfinkel2023deidentifying,
  title={De-Identifying Government Datasets: Techniques and Governance},
  author={Garfinkel, Simson and Guttman, Barbara and Near, Joseph and Dajani, Aref and Singer, Phyllis},
  institution={National Institute of Standards and Technology}, number={NIST SP 800-188},
  year={2023}, doi={10.6028/NIST.SP.800-188}}

@inproceedings{stadler2022synthetic,
  title={Synthetic Data---Anonymisation Groundhog Day},
  author={Stadler, Theresa and Oprisanu, Bristena and Troncoso, Carmela},
  booktitle={31st USENIX Security Symposium}, pages={1451--1468},
  publisher={USENIX Association}, year={2022},
  url={https://www.usenix.org/conference/usenixsecurity22/presentation/stadler}}

@techreport{vassilev2025adversarial,
  title={Adversarial Machine Learning: A Taxonomy and Terminology of Attacks and Mitigations},
  author={Vassilev, Apostol and Oprea, Alina and Fordyce, Alie and Anderson, Hyrum and Davies, Xander and Hamin, Maia},
  institution={National Institute of Standards and Technology}, number={NIST AI 100-2e2025},
  year={2025}, doi={10.6028/NIST.AI.100-2e2025}}

@inproceedings{anil2024manyshot,
  title={Many-Shot Jailbreaking},
  author={Anil, Cem and Durmus, Esin and Panickssery, Nina and Sharma, Mrinank and Benton, Joe and Kundu, Sandipan and Batson, Joshua and Tong, Meg and Mu, Jesse and Ford, Daniel and Mosconi, Francesco and Agrawal, Rajashree and Schaeffer, Rylan and Bashkansky, Naomi and Svenningsen, Samuel and Lambert, Mike and Radhakrishnan, Ansh and Denison, Carson and Hubinger, Evan J. and Bai, Yuntao and Bricken, Trenton and Maxwell, Timothy and Schiefer, Nicholas and Sully, James and Tamkin, Alex and Lanham, Tamera and Nguyen, Karina and Korbak, Tomasz and Kaplan, Jared and Ganguli, Deep and Bowman, Samuel R. and Perez, Ethan and Grosse, Roger Baker and Duvenaud, David},
  booktitle={Advances in Neural Information Processing Systems 37}, year={2024},
  doi={10.52202/079017-4121}}

@misc{perez2022redteaming,
  title={Red Teaming Language Models with Language Models},
  author={Perez, Ethan and Huang, Saffron and Song, Francis and Cai, Trevor and Ring, Roman and Aslanides, John and Glaese, Amelia and McAleese, Nat and Irving, Geoffrey},
  year={2022}, eprint={2202.03286}, archivePrefix={arXiv}, primaryClass={cs.CL},
  doi={10.48550/arXiv.2202.03286}, url={https://arxiv.org/abs/2202.03286}}

@misc{ganguli2022redteaming,
  title={Red Teaming Language Models to Reduce Harms: Methods, Scaling Behaviors, and Lessons Learned},
  author={Ganguli, Deep and Lovitt, Liane and Kernion, Jackson and Askell, Amanda and Bai, Yuntao and Kadavath, Saurav and Mann, Ben and Perez, Ethan and Schiefer, Nicholas and Ndousse, Kamal and Jones, Andy and Bowman, Samuel and Chen, Anna and Conerly, Tom and DasSarma, Nova and Drain, Dawn and Elhage, Nelson and El-Showk, Sheer and Fort, Stanislav and Hatfield-Dodds, Zac and Henighan, Tom and Hernandez, Danny and Hume, Tristan and Jacobson, Josh and Johnston, Scott and Kravec, Shauna and Olsson, Catherine and Ringer, Sam and Tran-Johnson, Eli and Amodei, Dario and Brown, Tom and Joseph, Nicholas and McCandlish, Sam and Olah, Chris and Kaplan, Jared and Clark, Jack},
  year={2022}, eprint={2209.07858}, archivePrefix={arXiv}, primaryClass={cs.CL},
  doi={10.48550/arXiv.2209.07858}, url={https://arxiv.org/abs/2209.07858}}

@inproceedings{mitchell2019modelcards,
  title={Model Cards for Model Reporting},
  author={Mitchell, Margaret and Wu, Simone and Zaldivar, Andrew and Barnes, Parker and Vasserman, Lucy and Hutchinson, Ben and Spitzer, Elena and Raji, Inioluwa Deborah and Gebru, Timnit},
  booktitle={Proceedings of the Conference on Fairness, Accountability, and Transparency},
  pages={220--229}, publisher={Association for Computing Machinery}, year={2019},
  doi={10.1145/3287560.3287596}}
\end{document}